\documentclass[useAMS,usenatbib]{mn2e}

\newcommand{\me}{m_{\rm e}}

\newcommand{\mec}{m_{\rm e} c}

\newcommand{\mecc}{m_{\rm e} c^2}

\newcommand{\sigmaT}{\sigma_{\rm T}}
\newcommand{\betaw}{\beta_{\rm w}}

\newcommand{\etaesc}{\eta_{\rm esc}}
\newcommand{\etasyn}{\eta_{\rm syn}}
\newcommand{\gammasyn}{\gamma_{\rm syn}}

\newcommand{\gammainj}{\gamma_{\rm inj}}

\newcommand{\gammacut}{\gamma_{\rm cut}}
\newcommand{\lambdamax}{\lambda_{\rm max}}
\newcommand{\Nb}{N_{\rm b}}
\newcommand{\nele}{n_{\rm e}}
\newcommand{\nph}{n_{\rm ph}}

\newcommand{\nusyn}{\nu_{\rm syn}}

\newcommand{\nuFnu}{\nu F_\nu}
\newcommand{\nuFnusyn}{(\nu F_\nu)_{\rm syn}}

\newcommand{\Resc}{R_{\rm esc}}

\newcommand{\tacc}{t_{\rm acc}}
\newcommand{\tsyn}{t_{\rm syn}}
\newcommand{\tesc}{t_{\rm esc,e}}
\newcommand{\ue}{u_{\rm e}}
\newcommand{\uB}{u_B}
\newcommand{\uph}{u_{\rm ph}}
\newcommand{\vw}{v_{\rm w}}

\newcommand{\zetalam}{\zeta \lambda_{\rm max}^{1-q}}
\usepackage{amsmath,amssymb}
\usepackage{bm}
\usepackage{graphicx}
\usepackage{ascmac}
\usepackage{cite}
\usepackage{mathrsfs}

\title[Stochastic Acceleration in TeV Blazars]{Synchrotron Self-Compton Emission by Relativistic Electrons under Stochastic Acceleration: Application to Mrk 421 and Mrk 501}
\author[J. Kakuwa et al.]
{Jun Kakuwa,$^{1}$\thanks{E-mail: kakuwa@theo.phys.sci.hiroshima-u.ac.jp (JK)}
Kenji Toma,$^{2,3}$
Katsuaki Asano,$^{4}$
Masaaki Kusunose,$^{5}$
\newauthor
and
Fumio Takahara$^{6}$
\\
$^{1}$Department of Physical Science, Hiroshima University, Higashi-hiroshima 739-8526, Japan\\
$^{2}$Astronomical Institute, Tohoku University, Sendai 980-8578, Japan\\
$^{3}$Frontier Research Institute for Interdisciplinary Sciences, Tohoku University, Sendai 980-8578, Japan\\
$^{4}$Institute for Cosmic Ray Research, The University of Tokyo, 5-1-5 Kashiwanoha, Kashiwa 277-8582, Japan\\
$^{5}$Department of Physics, School of Science and Technology, Kwansei Gakuin University, Sanda 669-1337, Japan\\
$^{6}$Department of Earth and Space Science, Osaka University, Toyonaka, 560-0043, Japan\\}

\begin{document}

\date{Accepted Y*** M******* D*. Received Y*** M******* D*; in original form Y*** M****** D*}
\pagerange{\pageref{firstpage}--\pageref{lastpage}} \pubyear{Y***}

\maketitle

\label{firstpage}

\begin{abstract}
We examine the applicability of the stochastic electron acceleration to two high synchrotron peaked blazars, Mrk 421 and Mrk 501, assuming synchrotron self-Compton emission of gamma-rays.
Our model considers an emitting region moving at relativistic speed, where non-thermal electrons are accelerated and attain a steady-state energy spectrum together with the photons they emit. 
The kinetic equations of the electrons and photons are solved numerically, given a stationary wave number spectrum of the magnetohydrodynamic (MHD) disturbances, which are responsible for the electron acceleration and escape. 
Our simple formulation appears to reproduce the two well-sampled, long-term averaged photon spectra. 
In order to fit the model to the emission component from the radio to the X-ray bands, we need both a steeper wave spectral index than the Kolmogorov spectrum and efficient particle escape. 
Although the model provides a natural explanation for the high-energy cutoff of the electron energy distribution, the derived physical parameters raise a problem with an energy budget if the MHD waves with the Alfv{\'e}n velocity are assumed to be the acceleration agent. 
\end{abstract}


\section{INTRODUCTION}
\label{sec:int}

Blazars, one of the classes of active galactic nuclei (AGNs), have highly collimated jets with relativistic speed pointing close to our line of sight \citep{Urry1995a, Marscher2009a}. 
Characteristics of X-ray variability of TeV emitting blazars \citep{Kataoka2008a} with time-scale of $\lesssim 1$~day measured in the observer frame suggest that the emitting region of the AGN jets that we see in blazars (so-called ``blazar region'') has the length scale of its structure extending up to $\sim 10^{16}$~cm in its comoving frame when it is estimated in this energy band. 
Since the electromagnetic radiation from that region is observed dominantly by virtue of the relativistic beaming effect, blazars are good candidates to investigate the energy conversion process of the AGN jets. 
Photon spectral energy distributions (SEDs) of blazars extend from the radio to the gamma-ray bands \citep{Fossati1998a} and are dominated by non-thermal emission. 
This demonstrates that non-thermal particles are accelerated in the blazar region. 

Blazars are categorized based on their SED peak frequency of the component that extends from the radio up to the X-ray bands in the $\nuFnu$ representation \citep{Abdo2010c}. 
For high synchrotron peaked blazars (HSPs), whose peak frequency is in the UV or X-ray band, the emission spectrum has been primarily interpreted as the synchrotron self-Compton (SSC) process: the synchrotron emission by the non-thermal relativistic electrons (and positrons) below the X-ray band and the inverse Compton (IC) scattering of these electrons off the synchrotron photons for the gamma-ray band \citep{Maraschi1994a, Sikora2001a, Dermer2012a}. 

In modeling observed SEDs, most of the studies assume either a broken power-law form with arbitrary indices for the energy distribution of the radiating non-thermal electrons \citep[e.g.,][]{Tavecchio2001a} or the energy distribution calculated by solving the kinetic equation under injection of a power-law spectrum, energy losses, and escape (or adiabatic cooling) near and downstream of a shock \citep[e.g.,][]{Mastichiadis1997a, Kusunose2000a}. 
Here we mention three implications on the shape of the electron energy distribution (EED) from such studies: 
(i) When a broken power-law EED is applied in the one-zone approximation, the change of the spectral index below and above the break energy tends to be larger than 1 \citep{Tagliaferri2008a, Tavecchio2010a, Zhang2012a}. 
This differs from a prediction of the simple one-zone model in which the break is caused by the competition between radiation cooling and escape. 
(ii) It is known that the broken power-law form of an EED is not always adequate to reproduce observed well-sampled SEDs. 
Purely phenomenologically, adding an additional break to the EED (i.e., two breaks in total) can improve the fit \citep{Kataoka1999a, Aharonian2009a, Abdo2011b, Abdo2011a}. 
The physical origin of these breaks (or bending of the EED) is not clear. 
(iii) If a population of non-thermal electrons form an SED from the radio band ($\sim 10^{10}$--$10^{12}$~Hz) to a higher frequency (e.g., the synchrotron peak), a hard EED is required. 
\citet{Kataoka2000a} reproduced the SED of this frequency range with the index of the EED of -1.35 for PKS 2155-304, and \citet{Kino2002a} -1.4, -1.6, and -1.8 for PKS 2155-304, Mrk 421, and Mrk 501, respectively. 
It has been shown that in principle such hard electron indices can be attained by the first-order Fermi acceleration in the test particle approximation, although the results are sensitive to conditions near the shock (e.g., inclination angle of the background magnetic field, the shock velocity, and the nature of particle scattering upstream and downstream of the shock) \citep[e.g.,][]{Bednarz1996a, Niemiec2004a, Lemoine2006a}. 
Note that since it is uncertain whether an EED of a single population is reflected in an SED from the radio band, it is usual not to intend to fit this range \citep[e.g.,][]{Krawczynski2004a, Kusunose2008a, Rani2013a}. 

There is also an implication on the acceleration time-scale of electrons: (iv) If the balance between synchrotron cooling and acceleration time determines the high-energy cutoff energy of an EED, 
the acceleration time of electrons with Lorentz factor $\gamma$, normalized by the gyroperiod of the same electrons, is given as
$\tau_{\rm acc}(\gammacut) = 3e/(\sigmaT B \gammacut^2)$, where $e$ is the charge of an electron, $\sigmaT$ the Thomson cross section, $B$ the magnetic field, and $\gammacut$ the electron Lorentz factor at the cutoff. 
Adopting $B = 0.1$~G and $\gammacut = 10^5$, which are derived for emission regions of some TeV blazars with the conventional one-zone SSC model \citep{Kino2002a}, one finds $\tau_{\rm acc} \sim 10^6$ \citep{Inoue1996a}. 
If it is assumed that only the first-order Fermi mechanism at a single shock works in the blazar region, both such a long time to accelerate the cutoff electrons and the hard spectra ((iii)) imply that conditions around the shock would presumably be very restricted, when extrapolated from the test particle simulations \citep[e.g.,][]{Bednarz1996a, Niemiec2004a, Lemoine2006a}. 
This issue may give an important implication on electron acceleration processes, though the spatial dependence of the magnetic field and the IC dominance on electron cooling can change the value of $\tau_{\rm acc}(\gammacut)$ to some extent and that one could not evaluate $\tau_{\rm acc}(\gammacut)$ well when only a weak constraint on the cutoff energy is obtained \citep[e.g.,][]{Abdo2011b, Abdo2011a}. 

Instead of interpreting the implications for the non-thermal EED mentioned above in terms of the first-order Fermi mechanism, stochastic acceleration (SA), which is described as diffusion in momentum space, can be invoked to be qualitatively consistent with them. 
Because the models including the SA, energy loss processes, and escape from a finite acceleration region can form various shapes of an EED \citep{Becker2006a, Stawarz2008a}, the SA has the possibility to account for the above features (i)--(iv) simultaneously. 
In addition, since the first-order Fermi mechanism necessitates scattering of particles, in the situation where it really acts slowly as mentioned in (iv), it may be accompanied by the SA effectively. 

The emission models including the SA have been applied to recent observations of blazars in many papers \citep[e.g.,][]{Katarzynski2006b, Massaro2006a, Ushio2010a, Weidinger2010a, Lefa2011b, Tramacere2011a, Yan2012a, Cao2013a}. 
They often focus on either limited portions of observed spectra (e.g., near the cutoffs of the synchrotron and the IC components) or relatively hard flaring spectra, and in some cases add the SA to power-law acceleration models or employ multi-zone models. 
We also made a time-dependent simulation with a radially structured jet model including the SA in \citet{Asano2014a}. 
In general, hard flaring spectra are compatible with a feature of the SA. 
A possibility of the SA for a slowly variable component is implied with the observation of Mrk 421 in the X-ray band by \citet{Ushio2009a}. 

In the present paper, we focus on the issues on the shape of EED, which have been raised mainly from one-zone steady-state models which assume a (broken) power-law EED. 
In order to investigate the validity of the SA to these issues, we apply the one-zone steady-state model employing a simple form of the SA and SSC emission to well-sampled long-term averaged emission of blazars. 
The observations compared with the model are the multiwavelength campaigns on Mrk 501 and 421, reported in \citet{Abdo2011b, Abdo2011a}. 

The paper is organized as follows. 
We describe our model in Section \ref{sec:Mod} and the procedure of application to the observations in Section \ref{sec:Fit}. 
The results are shown in Section \ref{sec:Res}, followed by the discussion in Section \ref{sec:Dis}. 
Section \ref{sec:Sum} is a summary. 

\section{MODEL} \label{sec:Mod}

We assume that a finite region where electrons are continuously injected into the SA (hereafter referred to as blob) is formed intermittently, propagates on a jet, and then reaches a steady state, forming the blazar region. 
We suppose that one of the blobs dominates an observed overall SED at a given time on average \citep{Sikora1997b}, and then calculate the synchrotron and the SSC emission from the blob by solving the kinetic equations of non-thermal electrons and photons inside the blob. 

The blob moves relativistically with Lorentz factor $\Gamma$ in the laboratory frame at an angle $\Gamma^{-1}$ to our line-of-sight. 
The Doppler factor of the blob is approximated as $\Gamma$. 
The shape of the blob is taken as cylindrical, characterized by the size longitudinal to the direction of its motion and that normal to it. 
Denoting them as $R_{\|}$ and $R_{\perp}$ respectively in the blob comoving frame, one can write the volume of the blob as 
\begin{equation}
 V = \pi R_{\perp}^{2} R_{\|} \, .\label{eq:V} 
\end{equation}
The two length scales relate to the formation of the blob in the jet. 
Neither acceleration/deceleration nor adiabatic change of the blob is considered. 

For simplicity, we assume that the accelerated electrons are always isotropically and homogeneously distributed in the blob frame and that the time evolution of their phase-space distribution is described as the pure momentum diffusion equation \citep{Hall1967a, Blandford1987a, Ostrowski1997a}. 
Taking some other effects into consideration, we calculate the steady-state EED by solving the kinetic equation in the blob frame given by \citep{Stawarz2008a}
\begin{align} 
\label{eq:kin_ele}
{\partial \nele(\gamma) \over \partial t} 
=& - {\partial \over \partial \gamma} \,\left[ \left( {2 \, D_{\rm e} \over \gamma} 
   + \dot{\gamma}_{\rm rad} \right) \, \nele(\gamma)
   - D_{\rm e}\, {\partial \nele(\gamma) \over \partial \gamma} \right] \nonumber \\[1ex]
 & - {\nele(\gamma) \over \tesc} 
   + q_{\rm inj}(\gamma) \, ,
\end{align}
where the accelerated electrons are relativistic enough that their momentum is approximated as $\gamma \mec$, $\me$ is the electron mass, $c$ the speed of light, $\nele(\gamma)$ the number density of the accelerated electrons per unit $\gamma$, $D_{\rm e}$ the diffusion coefficient of the electrons in $\gamma$ space due to the SA, $\dot{\gamma}_{\rm rad}$ the rate of the energy change due to the synchrotron and SSC processes, $\tesc$ the characteristic time for the electrons to escape from the blob, and $q_{\rm inj}(\gamma)$ the electron injection rate into the acceleration process per unit volume per unit $\gamma$. 
All the quantities in equations (\ref{eq:kin_ele}) and (\ref{eq:kin_ph}) below are defined in the blob comoving frame. 
The energy distribution of the photons emitted by the non-thermal electrons is isotropic and homogeneous in the blob frame. 
The photon kinetic equation is given by \citep{Li2000a}
\begin{equation}
\label{eq:kin_ph}
\frac{\partial \nph(\epsilon)}{\partial t} 
= \dot{n}_{\rm syn}(\epsilon)
+ \dot{n}_{\rm SSA}(\epsilon)
+ \dot{n}_{\rm IC}(\epsilon)
- \frac{\nph(\epsilon)}{t_{\rm esc,ph}} \, ,
\end{equation}
where $\epsilon$ is the photon energy in unit of $\mecc$, $\nph(\epsilon)$ the number density of photons per unit $\epsilon$, the first three terms on the right-hand side are the photon production and absorption rate by the electrons per unit volume per unit $\epsilon$, and $t_{\rm esc,ph}$ the photon escape time from the blob. 
The particular form of each term in equations (\ref{eq:kin_ele}) and (\ref{eq:kin_ph}) is described in the following. 

The diffusion of the electrons in $\gamma$ space can be caused by the scattering off magnetohydrodynamic (MHD) disturbances. 
Assuming that the disturbances are excited uniformly throughout the blob and are a superposition of small-amplitude waves with the phase speed (in unit of $c$) of $\betaw = \vw/c$, we can estimate the diffusion coefficient $D_{\rm e}$ for fast electrons as \citep{Longair1992a, Ostrowski1997a, Schlickeiser2002a} 
\begin{equation}
\label{eq:dif_ele}
D_{\rm e} 
\sim
\frac{\gamma^2}{r_g/v} 
\left(\frac{\vw}{v}\right)^2 
\left( \frac{\delta B_{\rm eff}}{B} \right)^2 \, ,
\end{equation}
where $v~(\gg \vw)$ is the electron speed ($\approx c$ at present), $r_g = r_g(\gamma) = \gamma \mecc/e B$ the gyroradius of the electrons, $B$ the root mean square of the magnetic field randomly oriented at scales larger than kinetic ones, $\delta B_{\rm eff}$ the disturbed magnetic field effectively contributing to the scattering of the electrons with Lorentz factor $\gamma$ with $\delta B_{\rm eff}/B$ being sufficiently less than unity, and we consider that the gyroresonance interaction between the disturbances and the electrons at the wave number $k \sim r_g^{-1}$. 
The factor $(r_g/v)^{-1} (\delta B_{\rm eff}/B)^2$ represents the pitch-angle diffusion coefficient under such situation \citep{Wentzel1974a}. 
In order to accelerate electrons to higher energies, a wave spectrum distributed in a wide $k$ range is needed. 
Assuming that the wave spectrum is stationary with a power-law spectrum $\delta B_k^2 \propto k^{-q}$, containing larger amount of energy at smaller wave number, that is, $q > 1$, we write $\delta B_{\rm eff}^2$ as 
\begin{equation}
\delta B_{\rm eff}^2 \sim \zeta B^2 \left( \frac{\lambdamax}{r_g} \right)^{1-q} \, ,\label{eq:Beff}
\end{equation}
where $\lambdamax$ is the longest wavelength of the wave spectrum and $\zeta$ the ratio of the energy density of the disturbed magnetic field to the mean magnetic field at the scale of $\lambda_{\rm max}$. 
By definition, $\zeta$ is less than unity. 
From equations (\ref{eq:dif_ele}) and (\ref{eq:Beff}), one sees $D_{\rm e} \propto \gamma^q$. 

The escape term describes spatial transport of the particles in the blob. 
For simplicity, we take the average length scale the particles propagate until escape, denoted by $\Resc$ in the blob frame, as ${\rm min}\{ R_\perp, R_\| \}$. 
We adopt the spatial diffusion time $\tesc = \tesc(\gamma) = \Resc^2/\kappa \propto \gamma^{q-2}$ as the electron escape time, where $\kappa = l c/3 = r_g c (B/\delta B_{\rm eff})^2/9$ is the spatial diffusion coefficient along the magnetic field, and $l \propto \gamma^{2-q}$ the mean-free path of the electrons \citep{Blandford1987a, Longair1992a}. 
Since treating the spatial transport as diffusion is improper for electrons with $l \gtrsim \Resc$ (, in particular for higher energy electrons if $q < 2$), we set $\tesc = \Resc/c$ for electrons with $l > \Resc$ in our calculation. 
As the photon escape time, we adopt the light-crossing time $\Resc/c$. 

The rate of the energy change of the accelerated electrons due to radiation can be written as $\dot{\gamma}_{\rm rad} = \dot{\gamma}_{\rm syn} + \dot{\gamma}_{\rm IC}$, where $\dot{\gamma}_{\rm syn}$ and $\dot{\gamma}_{\rm IC}$ are the rate due to the synchrotron and the IC emission, respectively. 
The corresponding photon source terms in Equation (\ref{eq:kin_ph}) are $\dot{n}_{\rm syn}$ and $\dot{n}_{\rm IC}$, respectively. 
Target photons of the IC scattering are only the photons originally emitted through the synchrotron process by the non-thermal electrons. 
Synchrotron self-absorption (SSA), which is important for the radio band spectra in the observer frame, is taken into consideration only as a sink term of photons, denoted by $\dot{n}_{\rm SSA}$. 
Electron-positron pair production/annihilation is not important for the parameter values we will choose and is also neglected. 

The synchrotron loss rate $\dot{\gamma}_{\rm syn}$, the corresponding photon production rate $\dot{n}_{\rm syn}(\epsilon)$, and the rate of SSA $\dot{n}_{\rm SSA}(\epsilon) = - c \, \alpha_{\rm SSA} \, \nph(\epsilon)$ are calculated following \citet{Rybicki1979a} and utilizing a numerical approximation given by \citet{Finke2008a}, where $\alpha_{\rm SSA}$ is the SSA coefficient. 
IC photon production rate $\dot{n}_{\rm IC}(\epsilon)$ \citep{Li2000a} and the corresponding energy change rate of the electrons $\dot{\gamma}_{\rm IC}$ are calculated with an approximate form of the IC scattering rate, following \citet{Jones1968a} and \citet{Blumenthal1970a}. 

Without treating thermal population of electrons, we just inject electrons to the acceleration process monoenergetically at a constant rate, $q_{\rm inj}(\gamma) \propto \delta(\gamma - \gammainj)$, where $\delta$ is the delta function, and $\gammainj$ the injection Lorentz factor fixed at 10 in this paper. 
Denoting the total electron injection luminosity as $L_{\rm inj}$, we relate $q_{\rm inj}$ with $L_{\rm inj}/\gammainj \mecc V$. 

An observed SED of photons radiated isotropically in the blob rest frame is calculated with the average photon escape rate $\epsilon \nph(\epsilon)V/t_{\rm esc,ph}$, in the $\nuFnu$ representation, by
\begin{equation}
 \nuFnu = \frac{\Gamma^4}{4 \pi d_{\rm L}^2} \frac{\epsilon^2 \mecc \nph(\epsilon) V}{t_{\rm esc,ph}} \, , \label{eq:nuFnu} 
\end{equation}
where $d_L$ is the luminosity distance, computed assuming the cosmological parameters of $H_0 = 71 \hbox{km}\, \hbox{s}^{-1} \hbox{Mpc}^{-1}$,  $\Omega_{\rm m0} = 0.27$, and $\Omega_{\Lambda 0} = 0.73$.
Photon energy in the observer frame relates to $\epsilon$ as $h \nu = \Gamma \epsilon \mecc / (1+z)$, where $h$ is the Planck constant, and $z$ the redshift. 
Steady-state energy spectra of the electrons and photons are computed numerically by solving equations (\ref{eq:kin_ele}) and (\ref{eq:kin_ph}) \citep{Press1992a, Park1996a}. 
We tested our numerical code with some analytical solutions \citep{Stawarz2008a, Mertsch2011a} and found that they are in good agreement. 
\section{FITTING PROCEDURE}
\label{sec:Fit}

To make comparison between observations and the model, we give eight physical parameters: $\Gamma$, $B$, $\Resc$, $V$, $L_{\rm inj}$, $\betaw$, $q$, and $\zetalam$, where the last one is included in equation (\ref{eq:Beff}).
Noting that $\lambdamax$ can not be constrained from photon observations, we do not consider the value of $\zeta$ or $\lambda_{\rm max}$ separately. 

The steady-state EED, except for its normalization, is determined by the balance among acceleration, cooling, and escape; when considering only a synchrotron component, one needs to pay attention to only three quantities which consist of some of the above parameters: $q$, synchrotron cooling efficiency $\etasyn = \etasyn(\gamma) = \tacc/\tsyn$, and electron escape efficiency $\etaesc = \etaesc(\gamma) = \tacc/\tesc$, where $\tacc = \tacc(\gamma) = \gamma^2/D_{\rm e} \propto \gamma^{2-q}$, $\tesc = \tesc(\gamma) \propto \gamma^{q-2}$, and $\tsyn = \gamma/\dot{\gamma}_{\rm syn} \propto \gamma^{-1}$  \citep{Stawarz2008a}. 

From equation (\ref{eq:kin_ele}), in the energy range where the radiation cooling can be neglected ($\etasyn \ll 1$), the acceleration forms a power-law EED of $\nele(\gamma) \propto \gamma^{1-q}$ at the steady state in the energy range of $\etaesc \ll 1$, while in the range of $\etaesc \gtrsim 1$, the escape gradually softens the EED and forms a non-power-law one if $q < 2$. 
Such effect of particle escape has the potential to be a natural explanation of the shape of EEDs implied by observations mentioned in Section \ref{sec:int} ((i)--(iii)). 
Since the synchrotron component extends from the radio to the X-ray bands for HSPs, in the one-zone model one needs to choose the parameter region making electrons with $\gamma$ sufficiently lower than $\gammasyn$ satisfy $\etaesc > 1$, where $\gammasyn$ is the Lorentz factor defined by $\etasyn = 1$. 
Namely, we consider the case that the electron escape effect yields a curved EED.
Note that the case of $q \geq 2$ is unsuitable for reproducing the well-sampled, gradually softening synchrotron spectra because from equation (\ref{eq:kin_ele}) a power-law EED with the index of $d\ln \nele(\gamma)/d\ln \gamma = 1/2-(9/4+\etaesc)^{1/2}$ is formed in the energy range of $\etasyn \ll 1$ for $q = 2$ \citep{Stawarz2008a}, where $\etaesc$ is constant, and because for $q > 2$ higher-energy electrons have shorter mean-free path, in other words, shorter $\tacc$ and longer $\tesc$. 
Hence we consider the range of $1 < q < 2$ coupled with one of the assumptions accompanying equation (\ref{eq:Beff}). 

\begin{figure}
\includegraphics[width=1.0\columnwidth]{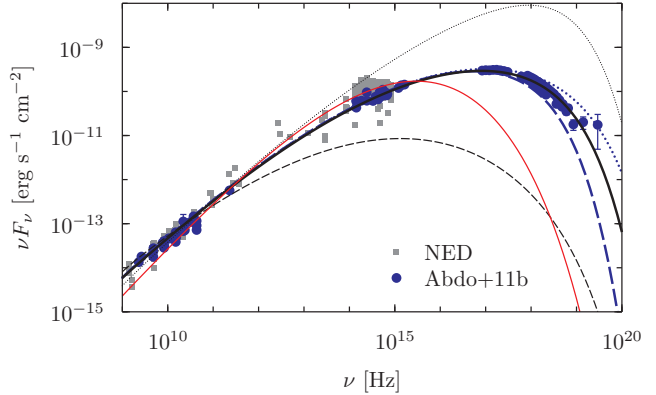}
\caption{Synchrotron spectra calculated for the different value of $q$, the different degree of escape efficiency $\etaesc(\gamma)$, and that of synchrotron cooling efficiency $\etasyn(\gamma)$. 
The black thick solid curve is for the same value of $q$, $\etaesc$ and $\etasyn$ as adopted in the left panel of Figure \ref{fig:SED}. 
The black thin dashed (dotted) curve represents the case of higher (lower) $\etaesc$ by a factor of 2, compared to the best-fitted one. 
The blue thick dashed (dotted) curve represents the case of higher (lower) $\etasyn$ by a factor of 2, compared to the best-fitted one. 
The red thin solid curve represents the case of $q=5/3$, which is varied from $q=1.85$ adopted for the best-fitted one. 
The plotted observational data are the synchrotron component of Mrk 421 from \citet{Abdo2011a} and NED.}
\label{fig:syn}
\end{figure}

\begin{figure*}
\includegraphics[width=1.0\linewidth]{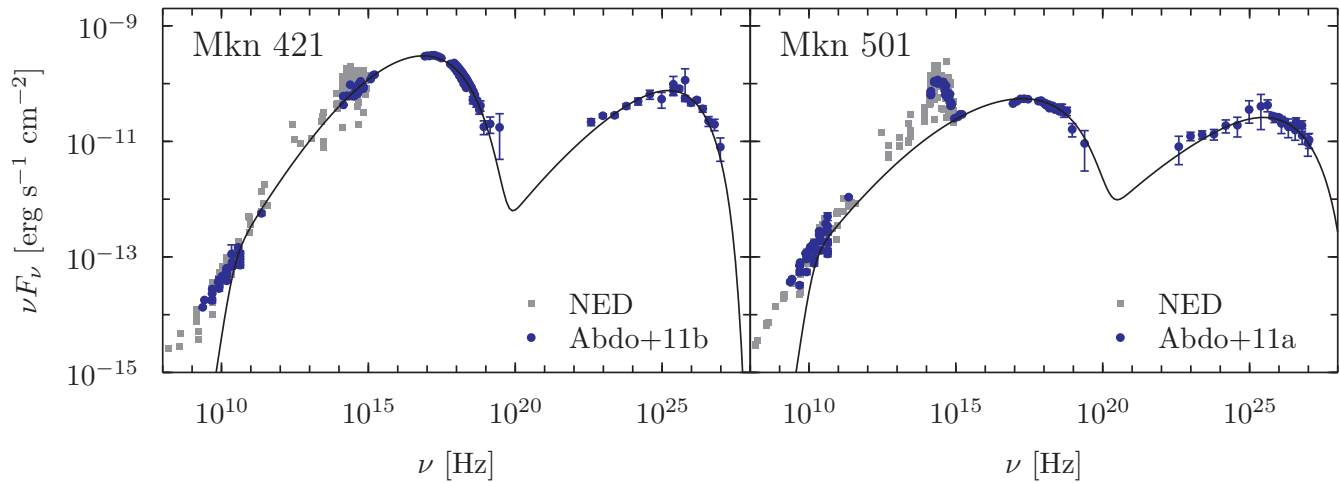}
\caption{Comparison between the model SEDs (solid curve) and the observational ones (blue circle). 
The left panel is for Mrk 421, and the right panel is for Mrk 501, where the observational data are taken from \citet{Abdo2011a} and \citet{Abdo2011b}, respectively. 
Non-simultaneous archival data taken from NED are plotted with the gray squares below $10^{15}$~Hz, for reference. }
\label{fig:SED}
\end{figure*}

We show in Figure \ref{fig:syn} with black curves the variation of the synchrotron spectrum with the degree of the electron escape efficiency. 
The black thick solid curve fits the observed spectrum of Mrk 421 with $q = 1.85$, $\etaesc(\gamma)$, and $\etasyn(\gamma)$ adopted in the left panel of Figure \ref{fig:SED} and displayed in the upper panel of Figure \ref{fig:timescale_421}. 
The plotted observational data are the synchrotron component of Mrk 421 (see \citet{Abdo2011a} for details). 
The black thin dashed (dotted) curve represents the case of higher (lower) $\etaesc(\gamma)$ by a factor of 2, compared to the best-fitted curve. 
The red thin solid curve in the same figure represents the case of $q=5/3$ with the value of $\etaesc(\gamma)$ for the electrons responsible for the synchrotron flux at $\approx 10^{14}$~Hz fixed. 
This curve ($q=5/3$) has stronger $\gamma$ dependence of the escape efficiency than the best-fitted curve ($q=1.85$) because $\etaesc \propto \gamma^{2(2-q)}$, so that the EED for the former makes a narrower spectral peak. 
As can be recognized from the black thin dashed and dotted curves, if one makes the escape effect less efficient for the red thin solid curve, the peak frequency becomes higher (up to which the synchrotron cooling becomes effective), and simultaneously the EED and so the SED become harder, and vice versa. 
This means that the value of $q$ needs to be closer to 2, that is, weaker $\gamma$ dependence of $\etaesc(\gamma)$ than at least the case of $q=5/3$ in order to obtain SEDs broad enough to fit with the observed ones. 
The adjustment of both the degree and the $\gamma$ dependence of the escape efficiency is needed to explain observed SEDs within this simple one-zone model. 

The blue thick dashed curve and dotted one in Figure \ref{fig:syn}, respectively, represent the case of higher and lower $\etasyn(\gamma)$ by a factor of 2, compared to the black thick solid one. 
These curves show that although the softening of the SEDs is caused by escape effect, it is insufficient to reproduce the high-energy cutoff, and radiative cooling is therefore required in this case. 

We adjust the model parameters to fit the model to the full observed SED including an SSC component, using a parameter set obtained from the fitting of the synchrotron component as the initial one and considering only the parameter region satisfying the following two conditions: (I) First, the characteristic synchrotron frequency of the electrons satisfying $l = \Resc$ is sufficiently high not to affect the observed SED. 
We consider only the case that the electrons with Lorentz factor 10 times larger than that corresponding to the maximum frequency of the observed synchrotron component satisfy $l < \Resc$. 
This choice means that we have the restriction $\betaw < \etaesc^{-1/2} < 1$ at this electron Lorentz factor because $\etaesc = (l/\betaw \Resc)^2$ by definition and at least $\etaesc > 1$ for electrons contributing to the flux at the frequencies around and above the synchrotron peak (as mentioned above in this section). 
(II) Second, since at least the value of $(\delta B_{\rm eff}/B)^2$ for the electrons treated in equation (\ref{eq:kin_ele}) which are responsible for observed emissions needs to be sufficiently low, we take $(\delta B_{\rm eff}/B)^2 < 0.2$ for the electrons with the Lorentz factor $10^2 \gammasyn$, which leads to $\zetalam < 0.2 (r_g(10^2 \gammasyn))^{1-q}$ from equation (\ref{eq:Beff}). 

\section{Results} 
\label{sec:Res}

\begin{table*}
\caption{Physical parameters used in the calculation shown in Figure \ref{fig:SED}. }
\label{tab:par}
\begin{center}
\small
\begin{tabular}{llllllllll}
\hline \hline
        & $\Gamma$ & $B$   & $\Resc$     & $V$              & $L_{\rm inj}$         & $\betaw$ & $q$  & $\zetalam$            & $\gammainj$ \\
unit    &          & G     & $10^{14}$cm & $10^{47}$cm$^3$  & $10^{39}$erg s$^{-1}$ &          &      & $10^{-14}$cm$^{1-q}$  &    \\
\hline
Mrk 421 & 35       & 0.081 & 3.5         & 1.9              & 1.6                   & 0.16     & 1.85 & 4.7                   & 10 \\
Mrk 501 & 31       & 0.014 & 28          & 95               & 15                    & 0.23     & 1.92 & 0.12                  & 10 \\
\end{tabular}
\end{center}
\end{table*}

We apply the model described in Section \ref{sec:Mod} to the well-sampled long-term averaged SEDs of two HSPs, Mrk 421 and Mrk 501. 
The results of the fits to the overall observed SEDs are shown in Figure \ref{fig:SED}. 
The adopted parameters are tabulated in Table \ref{tab:par}. 
The observational data for Mrk 421 was taken from 2009 January 19 (MJD 54850) to 2009 June 1 (MJD 54983) (Figure 8 in \citet{Abdo2011a}) and for Mrk 501 from 2009 March 15 (MJD 54905) to 2009 August 1 (MJD 55044) (Figure 8 and Figure 9 in \citet{Abdo2011b}). 
For the gamma-ray band of Mrk 501, we adopt the data plotted in Figure 9 of \citet{Abdo2011b} with filled circles, which do not include the time interval MJD 54952--54982 for the time averaging. 
The excluded term has significant flux variations in the gamma-ray band (e.g., a flux up to about five times higher than the average one at TeV energy) enough to cause a discrepancy between averaged {\it Fermi}-LAT data and non-flaring VERITAS data in overlapping energy range if included for the time averaging (see \citet{Abdo2011b} for more details). 
During the campaign Mrk 421 was in a low activity state at all wavebands (e.g., flux variations are typically less than a factor of two relative to the average flux level in gamma-ray band) \citep{Abdo2011a}. 
The data we adopted are suitable ones currently available for modelling the typical low state of these objects. 
The archival data plotted in Figure \ref{fig:SED} for reference was taken from NED. 

As shown in Figure \ref{fig:SED}, the two samples appear to be fitted well. 
Here we should note that the reduced chi-square values of the fits shown in Figure \ref{fig:SED} are $3.9$ and $1.3$, respectively, for Mrk 421 and Mrk 501 for the gamma-ray band ($3.8 \times 10^{22}$--$9.6 \times 10^{26}$~Hz for Mrk 421 and $3.9 \times 10^{22}$--$1.1 \times 10^{27}$~Hz for Mrk 501)\footnote{For the frequency range below the X-ray band, since the observed data show different flux levels at very close frequencies with small statistical errors, we do not evaluate chi-square values.}, though they are not minimum ones. 
The current work is not able to specify the causes of the statistical poorness of the fits. 
This may possibly indicate the need for non-trivial spatial structure of the emission region and/or additional emission components such as an external Compton component in order to reproduce the actual data more accurately. 
As an example for the latter, the comptonization of an external radio photon field has been proposed so as to explain somewhat but systematically underestimated flux near $10^{23}$~Hz (see Section 4.2 in \citet{Asano2014a}). 
Nevertheless, we consider that the fits are sufficient for investigating the validity of SA to the issues described in Section \ref{sec:int}.

\begin{figure}
\includegraphics[width=1.0\columnwidth]{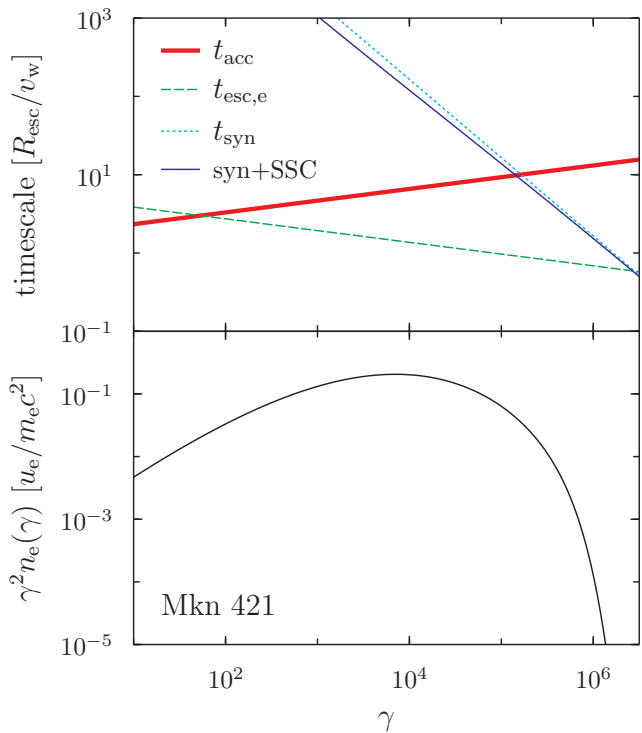}
\caption{Upper panel: The characteristic times of acceleration, escape and cooling for non-thermal electrons, corresponding to the left panel of Figure \ref{fig:SED}. 
The red thick solid line represents the acceleration time $\tacc$, the green dashed line the electron escape time $\tesc$, the blue thin solid curve the radiation cooling time at the steady state, $\gamma/\left| \dot{\gamma}_{\rm rad} \right|$, and the cyan dotted line the synchrotron cooling time $\tsyn$. 
Lower panel: Electron energy distribution of the non-thermal electrons, multiplied by $\gamma^2$. 
The electrons are injected at $\gamma = 10$ continuously at a constant rate. }
\label{fig:timescale_421}
\end{figure}

The characteristic synchrotron frequency of the electrons with the Lorentz factor $\gammainj$ is about $10^9$~Hz for Mrk 421, and about $10^8$~Hz for Mrk 501, measured in the observer frame. 
The break seen near $10^{10}$~Hz in the both panels in this figure is caused by SSA. 
In the upper panel of Figure \ref{fig:timescale_421}, the characteristic times of acceleration, escape, and cooling are displayed as a function of $\gamma$ for the parameters adopted for Mrk 421. 
The lower panel shows the calculated EED of Mrk 421. 
It can be recognized from this figure that gradual softening from the radio band to higher frequency is formed by an increasing dominance of escape over acceleration with an increasing electron energy and that the cutoff at the synchrotron peak is caused by the balance between acceleration and cooling time under rapid escape. 
Note that not $\gamma^2 \nele(\gamma)$ but $\gamma^3 \nele(\gamma)$ mimics the synchrotron spectrum \citep{Tramacere2011a}. 
Not only the softening by the efficient escape, but also cooling is required to well reproduce the observed shape of the cutoff in this case. 

The synchrotron component of Mrk 501 has a softer and broader portion, compared to Mrk 421, as seen in Figure \ref{fig:SED}. 
If one interprets such an SED from the radio band in the one-zone model, $q$ closer to 2 (adopted value is 1.92) is needed as mentioned in Section \ref{sec:Fit}. 
Treating the slope of the wave spectrum as a free parameter is essential to apply our simple model to various objects. 
At least the phenomenologies such as Kolmogorov-type $q=5/3$ and Kraichnan-type $q=3/2$ \citep{Goldstein1995a} are not fit for the application here; larger $q$ but smaller than 2 is needed.
We omit the figures for the time-scales and EED for Mrk 501 because it is basically similar to that for Mrk 421. 

The adopted parameter values are listed in Table \ref{tab:par}. 
We call them and the ones obtained with them the canonical values. 
By examining the relation between a model SED and physical quantities, here we show that some of them are constrained to the canonical values with the SED fitting while others are not. 
First, we note that the synchrotron component is not modified except by the influence of the IC cooling when we change the parameter values with the flux normalization of the synchrotron component and the values of $\etaesc$ and $\etasyn$ fixed as a function of synchrotron frequency measured in the observer frame, $\nusyn$, with the corresponding electron Lorentz factor $\gamma$, i.e., we obtain
\begin{eqnarray}
\nusyn^* &=& 1 = (B \gamma^2 \Gamma)^* \, ,\label{eq:nusyn} \\
\nuFnusyn^* &=& 1 = (L_{\rm syn} \Gamma^4)^* = (B^2 \gamma \ue \Gamma^4 V)^* \, ,\label{eq:nuFnusyn} \\
\etasyn^* &=& 1 = (B^q \gamma^{3-q} \betaw^{-2} (\zetalam)^{-1})^* \, , \\
(\etaesc^{1/2}/\etasyn)^* &=& 1 = (\betaw \Resc^{-1} B^{-2} \gamma^{-1})^* \, , 
\end{eqnarray}
where the ratio of a quantity to the canonical value is denoted by the superscript $^*$, $\nuFnusyn^*$ is the ratio of the flux normalization of the synchrotron component, $L_{\rm syn}$ the synchrotron luminosity measured in the blob frame, $\uB = B^2/8 \pi$ the energy density of the mean magnetic field, $\ue$ that of the non-thermal electrons, and $L_{\rm syn}^* = (\uB \gamma \ue V)^*$. 
Next, for the fixed synchrotron component, the IC component is not modified when:
\begin{eqnarray}
(u_{\rm syn}/\uB)^* &=& 1 = (\gamma \ue \Resc)^* \, , \label{eq:usynuB} \\
\gamma^* &=& 1 \, , \label{eq:gamma} \\
\Gamma^* &=& 1 \, , \label{eq:nusscmax} 
\end{eqnarray}
where $u_{\rm syn}$ is the energy density of the synchrotron photons, measured in the blob frame, and $L_{\rm syn}^* = (u_{\rm syn} V/\Resc)^*$. 
Note that the frequency range of the Thomson scattering regime of the IC component relative to the synchrotron component and the IC cutoff frequency roughly scale as $\gamma^{* 2}$ and $(\gamma \Gamma)^*$, respectively. 
From equations (\ref{eq:nusyn})--(\ref{eq:nusscmax}), we obtain
\begin{eqnarray}
B^* &=& \gamma^* = \Gamma^* = 1 \, , \\ 
\ue^* &=& V^{* -1} \, ,\label{eq:ue} \\
\Resc^* &=& V^* \, , \label{eq:Resc} \\
\betaw^* &=& V^* \, , \label{eq:betw} \\
(\zetalam)^* &=& V^{* -2} \, . \label{eq:zetlam}
\end{eqnarray}
Thus, $B$, $\Gamma$, and the total electron energy $\ue V$ are well constrained to the canonical values. 
Moreover, the SED is unchanged if the physical parameters of the blob are changed according to equations (\ref{eq:ue})--(\ref{eq:zetlam}) with $V^*$. 
That is, a parameter set adopted for an observed SED as shown in Figure \ref{fig:SED} includes only one degree of freedom and can vary the parameter values within the restrictions (I) and (II) in Section \ref{sec:Fit}. 
Although $L_{\rm inj}$, one of the fitting parameters, does not appear in the above analysis explicitly, one can rewrite equation (\ref{eq:ue}) in terms of $L_{\rm inj}$. 
Note that all the presented figures are independent of the uncertainty and our general conclusions are unaffected. 

For concreteness, we have shown in Table \ref{tab:par} the parameter values of $\Resc$, $V$, $L_{\rm inj}$, and $\zetalam$ by adopting the maximum allowed value of $\betaw$ from the restriction (I) described in Section \ref{sec:Fit}. 
The shown parameter sets can be scaled by $V^*$ in the range of $0.04 \leq V^* \leq 1$ for Mrk 421 and $0.09 \leq V^* \leq 1$ for Mrk 501 with the SEDs fixed, where the upper and lower bound come from (I) and (II) in Section \ref{sec:Fit}, respectively. 

\begin{table}
\caption{Energy contents in the blob for the parameters shown in Table \ref{tab:par}: the energy density of magnetic field, $\uB$; that of photons, $\uph$; and that of non-thermal electrons, $\ue$. }
\label{tab:ene}
\begin{center}
\small
\begin{tabular}{llllll}
\hline \hline
        & $\uB$                   & $\uph$                  & $\ue$                    \\
unit    & 10$^{-2}$ erg cm$^{-3}$ & 10$^{-2}$ erg cm$^{-3}$ & 10$^{-2}$ erg cm$^{-3}$  \\
\hline
Mrk 421 & 0.03                    & 0.02                    & 9                        \\
Mrk 501 & 0.0007                  & 0.002                   & 2                        \\
\end{tabular}
\end{center}
\end{table}

Energy densities of respective components, measured in the blob frame, are listed in Table \ref{tab:ene}. 
Although the magnitude relation between $\ue$ and $\uB$ is consistent with the conventional SSC model \citep[e.g.,][]{Kino2002a}, the non-thermal electrons possess much greater energy than the magnetic field, $\ue/\uB \gg 1$, compared to the previous results. 
This difference originates from relatively small $\Resc$, that is, the importance of the escape effect as seen from equations (\ref{eq:usynuB}) and (\ref{eq:gamma}). 
Within the uncertainty of the parameters, we can only make $\ue$ (and $\ue/\uB$) larger from the canonical values since $(\ue V)^* = 1$ is required to fix the SEDs from equation (\ref{eq:ue}), where $V^* \leq 1$. 

\section{DISCUSSION} 
\label{sec:Dis}

\subsection{Shape and the Number of the Blobs}
\label{subsec:geometry}
As seen in Table \ref{tab:par}, a relatively small $\Resc (\ll (V/\pi)^{1/3})$ is required in order to make the escape effect work to form the gradually softening SEDs. 
This means that the emission region is either disk-like or spindle like.

However, the extreme shape of the blob may be alleviated if one assumes that a number of the blobs with the same properties simultaneously contribute to an observed spectrum. 
The number of the blobs $\Nb$ can be easily included into the analysis in Section \ref{sec:Res} by multiplying $L_{\rm syn}$ in equation (\ref{eq:nuFnusyn}) by $\Nb$. 
Then $V^*$ in equations (\ref{eq:ue}) -- (\ref{eq:zetlam}) is replaced with $(V \Nb)^*$. 
It can be recognized from equation (\ref{eq:V}) and the equation modified from equation (\ref{eq:Resc}) with this replacement, $\Resc^* = (V \Nb)^*$, that for example, when increasing $\Nb$ and decreasing $V$ with $(V \Nb)^* = 1$, one can scale the ratio $\Resc/{\rm max}\{ R_\perp, R_\| \} = {\rm min}\{ R_\perp, R_\| \}/{\rm max}\{ R_\perp, R_\| \}$ up to unity as $(V \Nb^{3/2})^* = \Nb^{1/2}$ if $R_\perp > R_\| = \Resc$ or as $(V^2 \Nb^3)^* = \Nb$ if $R_\| > R_\perp = \Resc$ without changing both the SED and the parameter values (except for $L_{\rm inj}$). 
This is the case of multiple blobs contributing to an SED. 
In the above example, we have $\Resc/{\rm max}\{ R_\perp, R_\| \} < 1$ with $\Nb \lesssim 10^3$ and $\lesssim 10^2$ for Mrk 421 and Mrk 501, respectively. 
However, a large number which leads to $R_\perp \sim R_\|$ is rather unlikely in the spirit of the simple one-zone model approach.

\subsection{Energy Budget}
\label{subsec:energetics}

\begin{figure}
\includegraphics[width=1.0\columnwidth]{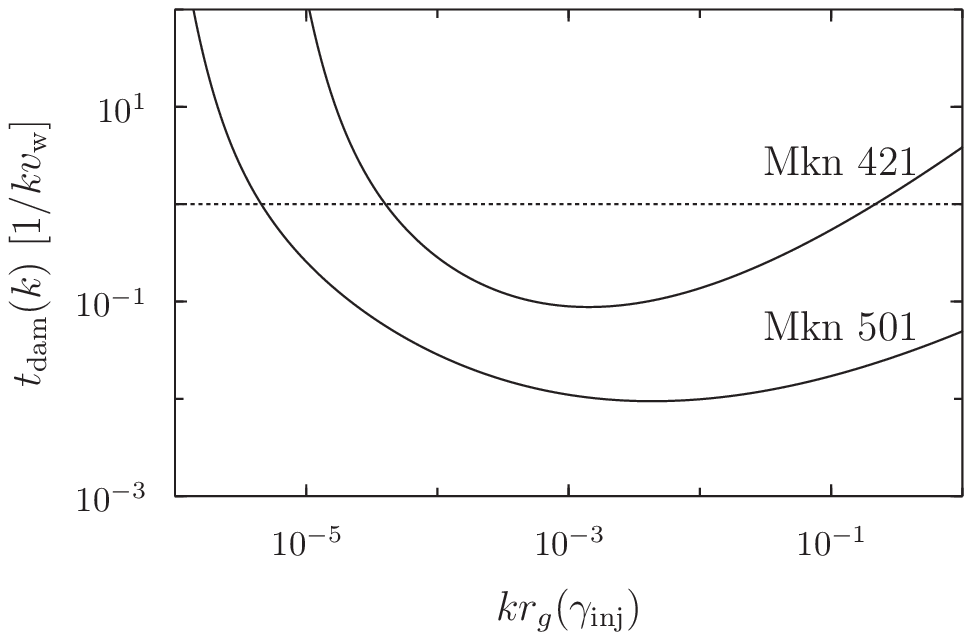}
\caption{The time-scale of damping of MHD disturbances through acceleration of non-thermal electrons in unit of $1/k \vw$ for the case of $u_{\rm w}(k) \sim \delta B_{\rm eff}^2/8 \pi$. 
The horizontal axis represents wave number $k$ normalized by the inverse of the gyroradius of the electrons with the injection Lorentz factor $\gammainj$. 
The dotted line indicates a lower limit of the wave cascading time. }
\label{fig:damp}
\end{figure}

Let us consider the energy balance between the waves and the accelerated particles. 
Since we have supposed the net energy transfer between the waves and the non-thermal electrons is from the former to the latter, the energy supply from large scale disturbances to small scale ones has to be rapid enough to meet the assumption of the stationary wave spectrum. 
We estimate the time-scale of damping of the wave energy caused by the electron acceleration, as a function of $k$, as 
\begin{equation}
 t_{\rm dam}(k) \sim \frac{u_{\rm w}(k)}{\dot{\mathcal{U}}_{\rm acc}} \, , 
\end{equation}
measured in the blob frame, where $u_{\rm w}(k)$ is the total energy density of wave modes around $k$, $\dot{\mathcal{U}}_{\rm acc} \equiv \gamma^2 \mecc \nele(\gamma)/\tacc(\gamma)$, and $k \sim r_g^{-1}$. 
If this is too short, waves would damp through particle acceleration and then $u_{\rm w}(k)$ would be modified. 
For illustrative purpose we display $t_{\rm dam}(k)$ in Figure \ref{fig:damp} for the steady state we have obtained, supposing $u_{\rm w}(k) \sim \delta B_{\rm eff}^2/8 \pi$. 
By comparing $t_{\rm dam}(k)$ to $1/k \vw$, which represents a lower limit of the time-scale of the wave cascading, we find that $t_{\rm dam}$ is smaller than $1/k \vw$ in an intermediate range of $k$. 
This means that before the blob reaching the steady state, the assumed wave spectrum would be modified and evolve toward a different equilibrium state.
The model is hence inconsistent in this case. 

Under our treatment of the electron transport, this encourages us to interpret $\vw$ as the sound speed (not as the Alfv{\'e}n velocity) since the wave energy can be much larger than the magnetic part in this case if the plasma beta of the emitting region is much larger than 1. 
It seems qualitatively reasonable because the energy of the magnetic field is much less than that of the particles in the blob. 
This implies that the fast mode waves may be the acceleration agent. 
(In this case $u_{\rm w}(k)$ and so $t_{\rm dam}(k)$ become larger in proportion to the plasma beta although the cascading time is also affected.)
The parameter sets shown in Table \ref{tab:par} have the value of $\vw$ close to the upper limit of the sound speed (i.e., $c/\sqrt{3}$), which seems likely in the blazar region. 
Anyhow, more sophisticated turbulence modeling accounting for particle acceleration is needed for self-consistency, which is beyond the scope of this paper. 

\section{Summary} 
\label{sec:Sum}

Motivated by the previous implications about the energy distribution of the non-thermal electrons in the blazar region ((i)--(iv) described in Section \ref{sec:int}), we have examined the stochastic acceleration by a weak random electromagnetic field as the acceleration mechanism of the electrons by modelling the SEDs of two HSPs, Mrk 421 and Mrk 501, from the radio to the gamma-ray bands. 
All the model parameters are time independent (Section \ref{sec:Mod}). 
The long-term time averaged SEDs appear to be well fitted by the emitting blob reaching a steady state (Figure \ref{fig:SED}). 
The blob can be considered to represent the jet emission region(s) mainly contributing to the observed SED on average in the long term. 
The goodness of the fits in gamma-ray band is, however, statistically not sufficient as mentioned in Section \ref{sec:Res} and more refined modelling are needed to reproduce the actual data more accurately. 
With the current simple model, we have obtained the following results. 

The model can naturally produce the broad and curved electron energy distribution, which have been suggested based on the conventional models so far (Section \ref{sec:int}). 
To realize such electron energy distribution, two points are crucial in the model: a stationary, power-law spectrum of MHD waves with a steeper spectral index than the Kolmogorov spectrum at the relevant wavelengths and efficient particle escape from the acceleration region (Figures \ref{fig:syn} and \ref{fig:timescale_421}). 
The former is associated with the latter through weaker energy dependence of the escape efficiency. 
As for the formation of the high-energy cutoff of the synchrotron component, radiative cooling becomes important. 

These results lead to a few issues with the obtained parameter values. 
First, the shape of the blob is very asymmetric having a relatively small side length compared with the volume of the blob to realize an efficient escape. 
Second, since emitted photons also efficiently escape, in order to have a reasonable value for the ratio of the energy density of the synchrotron photons to that of the magnetic field, which is strictly constrained by an observed spectrum, our model requires a large ratio of the energy density of the non-thermal electrons to that of the magnetic field (equation (\ref{eq:usynuB})), which ratio is actually orders of magnitude larger than the ones hitherto adopted for the two HSPs (Section \ref{sec:Res}).

A trivial but important issue of the model is whether and how such properties phenomenologically required for the emission region can be formed in the blazar jets. 
If we estimate the energy density of the MHD waves responsible for the electron acceleration on the assumption that they propagate at the Alfv{\'e}n velocity, the derived physical parameters raise a problem with an energy budget of MHD turbulence, i.e., the turbulence would be damped effectively by the particle acceleration (Section \ref{subsec:energetics}). 
The model is inconsistent in this case. 
The difficulty may be reduced to some extent if we assume that the waves propagate at the speed of sound in a high-beta plasma (i.e., the fast mode waves). 

Recently, \citet{Asano2014a} considered the emission from the electrons stochastically accelerated in the continuous and radially-structured jet, where the escape effect can be negligible. 
In \citet{Asano2014a} assuming the Kolmogorov spectrum ($q=5/3$) for the turbulent magnetic field, which is different from our treatment of $q$, they investigated the effect of the radial evolution of the physical parameters on the electron energy distribution and observed SED. 
They have shown that electron spectra become soft as required even when $q = 5/3$ by assuming the evolution of the electron injection rate, that of the momentum diffusion coefficient, and adiabatic cooling. 
Although their jet model is quite a contrast to the blob model in this paper, it is also able to form the curved electron energy distribution that can explain the same SED of Mrk 421 by the stochastic acceleration. 
Thus, although the stochastic acceleration is a viable model, the way to realize a broad but a non-power law  electron spectrum is not unique and further work is needed to solve the problem. 

\section*{Acknowledgments}

The authors thank D. Paneque for giving us the $\nuFnu$ data shown in \citet{Abdo2011b, Abdo2011a}. 
This research has made use of the NASA/IPAC Extragalactic Database (NED) operated by the Jet Propulsion Laboratory, California Institute of Technology, under contract with the National Aeronautics and Space Administration. 


\bibliographystyle{mn2e}
\bibliography{ref}

\bsp 


\label{lastpage}

\end{document}